\documentstyle[11pt]{article}

\begin{document}

\def\mxth{\mathsurround=0pt }
\def\xversim#1#2{\lower2.pt\vbox{\baselineskip0pt \lineskip-.5pt
  \ialign{$\mxth#1\hfil##\hfil$\crcr#2\crcr\sim\crcr}}}
\def\simgr{\mathrel{\mathpalette\xversim >}}
\def\simle{\mathrel{\mathpalette\xversim <}}
\newcommand{\GeV}{\mbox{ GeV}}
\newcommand{\sth}{$s^{2}(M_{Z})$}
\newcommand{\sths}{$s^{2}(M_{Z})$ }
\newcommand{\als}{$\alpha_{s}(M_{Z})$}
\newcommand{\alss}{$\alpha_{s}(M_{Z})$ }
\newcommand{\mhiggs}{$m_{h^{0}}$}
\newcommand{\mhs}{$m_{h^{0}}$ }

\title{\small PRECISION DATA PARAMETERS\\
AND GRAND UNIFICATION PREDICTIONS}

\author{\small Nir Polonsky\footnote{}\\
{\it Department of Physics, University of Pennsylvania,}\\
{\it Philadelphia, Pennsylvania,  19104, USA} \\}
\date{UPR-0588T}
\setlength{\baselineskip}{2.6ex}
\maketitle
\begin{abstract}
When realizing the Minimal Supersymmetric Standard Model (MS-SM)
within a (simple) Grand Unified Theory (GUT), the predicted ranges
for the strong coupling and the ratio of two Higgs doublet expectation
values are strongly correlated with the
weak angle and the $t$-quark mass. (The latter are
extracted from precision data.) We spell out those relations and point out
the implication for the mass of the light MSSM (CP even) Higgs boson.
\end{abstract}

When simultaneously considering coupling constant
unification within the
MSSM \cite{nath}; analysis of electroweak precision data
\cite{us} (which recently implied the consistency of the former with the data
\cite{firsts}); and a $SU_{5}$ (or similar) symmetry that relates the
$b$-quark and the $\tau$-lepton
Yukawa couplings (i.e., $h_{b} = h_{\tau}$ at the
unification point\footnote{
This holds if matter couples only to Higgs fields
in the fundamental representations -- `` a simple GUT''.
One usually assumes that some perturbation modifies the
coupling or the masses of the two light families where, in principle,
similar relations should, but do not, hold \cite{hall}.} \cite{mbmtau});
then the weak angle, \sths (as constrained by precision data);
the strong coupling, \alss (as predicted by coupling constant unification);
the $t$-quark (pole) mass, $m_{t}$ (as constrained by precision data,
by the relation $h_{b} = h_{\tau}$, and by requiring
a broken $SU_{2} \times U_{1}$); and the allowed range for
the ratio of the the two Higgs
doublet expectation values, $\tan\beta \equiv {\nu_{up}/ \nu_{down}}$
(as constrained by the relation $h_{b} = h_{\tau}$, and by requiring
a broken $SU_{2} \times U_{1}$); are all strongly correlated.
In turn, this implies a constrained scenario, and in particular
$m_{h^{0}} \simle 100$ ($110$) GeV
[for $m_{t} \simle 160$ ($170$) GeV] for the mass of the
light MSSM (CP even) Higgs boson.
For a complete discussion, we refer the reader to
Ref. 2, 6.  Similar issues were also studied in Ref. 7, 8.

The $\rho$-parameter relates \sths and $m_{t}$, i.e.,
(in the $\overline{MS}$ scheme)
\begin{equation}
s^{2} = s^{2}_{0} -
\left[ 1 - 2\alpha_{s}(m_{t})\frac{\pi^{2} + 3}{9\pi}\right]
\frac{3G_{F}}{8\sqrt{2}\pi^{2}}s^{2}_{0}\frac{1 -
s^{2}_{0}}{1 - 2s^{2}_{0}}[m_{t}^{2} - {m_{t}}_{0}^{2}]\;,
\label{eq1}
\end{equation}
where $s^{2}$ is $[s^{2}(M_{Z})](m_{t})$
and $s^{2}_{0} = s^{2}({m_{t}}_{0})$.
The $t$-quark (pole) mass
is independently constrained by the measured value of the
$ Z \rightarrow b\overline{b}$ vertex.
A two-parameter fit (allowing \mhs to vary from
$50 - 150$ GeV with a central value $m_{h^{0}} = M_{Z}$) to all $Z$, $W$,
and neutral-current data yields \cite{us}
\begin{equation}
s^{2}(M_{Z}) = 0.2326 \pm 0.0003 - 0.92\times 10^{-7}\GeV^{-2}
[m_{t}^{2} - (134\GeV)^{2}]\;,
\label{eq2}
\end{equation}
\begin{equation}
m_{t} = 134^{+23}_{-28} + 12.5\ln{\frac{m_{h^{0}}}{M_{Z}}}\;,
\label{eq3}
\end{equation}
where the $\pm 0.0003$ uncertainty
in (\ref{eq2}) is from logarithmic dependences
on \mhs and on $m_{t}$, while the leading quadratic dependence on $m_{t}$
in (\ref{eq2}) and the logarithmic dependence on \mhs in (\ref{eq3})
are explicitly displayed. (Note that when changing the central
value of \mhiggs, the central value of $m_{t}$ will change
in a correlated manner.)

The above is exact in the heavy MSSM limit, which is applicable in most
of the MSSM parameter space.
In that limit one has a SM-like light Higgs boson
and all other Higgs and supersymmetric particles
contribute negligibly to the electroweak observables.
Supersymmetric and Higgs particles
contributing non-negligibly to the $\rho$-parameter
will reduce, in general, the fitted value of $m_{t}$.
The same is  usually true for contributions to the
$ Z \rightarrow b\overline{b}$ vertex, unless we are in that region of
parameter space where, e.g., both, a chargino and a scalar-top, are light
enough \cite{jens}.

The prediction of the strong coupling reads (using $\alpha(M_{Z})^{-1}
= 127.9 \pm 0.1$)
\begin{equation}
\alpha_{s}(M_{Z}) = 0.124 \pm 0.001 \pm 0.008 + H_{\alpha_{s}}
+ 3.2 \times 10^{-7}\GeV^{-2}[m_{t}^{2} - (134\GeV)^{2}]\;,
\label{eq4}
\end{equation}
where the $\pm 0.008$ ($\pm 0.001$)  uncertainty in (\ref{eq4})
is theoretical (due to the input parameter error bars).
The former is due to the unknown values of the matching conditions
(i.e., threshold and nonrenormalizable effects)
at both scales, which are expected to be non-trivial. We discuss
these issues in detail in Ref. 2,
where we parametrize and estimate those effects.
The function $H_{\alpha_{s}}$
is negative but negligible, unless some Yukawa couplings are $ \simgr 1$,
where we obtain $H_{\alpha_{s}} \approx - (0.001 - 0.003)$.

Eq. (\ref{eq4}) and requiring $m_{b}(5 \GeV) \leq 4.45$ GeV (where
$m_{b}$ is the $\overline{MS}$ $b$-quark running mass) allow
only two regions in the $m_{t} - \tan\beta$ plane when imposing
the $SU_{5}$ relation $h_{b} = h_{\tau}$. This is illustrated in Fig. 1.
Non-trivial matching corrections can modify that relation by $\simle 15 \%$
(see Ref. 2), an effect that is accounted for in Fig. 1
(i.e., $m_{b}^{0}$, which is calculated numerically assuming naive
matching conditions, is corrected by a factor of $1 \pm 0.15$).
In order for the rescaling of the mass parameters
between the unification\footnote{We assume that to a good approximation
all masses are given near the unification scale
by a small number of universal parameters \cite{nath}.}
and weak scales to be consistent
with a broken $SU_{2}\times U_{1}$ at the weak scale, we have to further
constrain the $m_{t} - \tan\beta$ plane by requiring
$\tan\beta \geq 1$ and $h_{t} \geq h_{b}$ for the $t$ and $b$-quark
Yukawa couplings \cite{drees}.
Thus, within one standard deviation of the fit,
\begin{equation}
140 \simle m_{t} \simle 160 \GeV,
\label{eq5}
\end{equation}
and
\begin{equation}
\tan\beta = 0.935 + 1.68x -0.865x^{2} +16.62x^{3} - 16.06x^{4}
\pm \delta_{\beta}\;,
\label{eq6}
\end{equation}
where $x \equiv [(m_{t} - 140 \GeV) / (140 \GeV)]$ and $\delta_{\beta} \approx
0.05$ ($0.06$) for $m_{t} \simle 160$ GeV
($160 \simle m_{t} \simle 180$ GeV). The above formula
describes the low-$\tan\beta$ branch for
$140 \simle m_{t} \simle 180$ GeV.
Within two standard deviations (e.g., $m_{t} \approx 180$ GeV),
a solution with a very large $\tan\beta$
(s.t. $h_{t} \simgr h_{b} \simgr 1$) is allowed.
Such scenarios (that are compatible with a $SO_{10}$
symmetry) are studied in Ref. 11.

However, if $m_{t} \simle 165$ GeV, then $\tan\beta \approx 1$
and the light (CP even) Higgs boson mass will be determined
by the magnitude of the loop correction, $\Delta_{h^{0}}$.
The tree level mass term, $m_{h^{0}}^{T} \leq |M_{Z}\cos 2\beta|$,
vanishes in this limit. This was recently pointed out in Ref. 7.
We are currently studying the implications
of such a scenario to the full MSSM parameter space \cite{us2}.
We find that due to a possible large mixing in the scalar-top
sector $\Delta_{h^{0}}$ may be maximized, and thus \mhs can be
(given the above assumptions) slightly above $M_{Z}$.

To summarize, we pointed out that an attractive set of ideas --
 minimal grand unified supersymmetric standard model (as well
as some simple extensions) -- implies a very constrained
parameter space. Note that aside from $\tan\beta \geq 1$
we did not assume any specific structure of the mass parameters
in the theory. The effects of these were contained in a set of correction
functions that we estimated \cite{us}
and found to only slightly affect the results.
The effect of the above constraints on the MSSM mass parameters is
currently under study \cite{us2}.

\subsection*{Acknowledgments}
This work was done with Paul Langacker, and
was supported by the US Department of Energy
Grant No. DE-AC02-76-ERO-3071. It is pleasure to thank J. Erler for
useful discussions. I would like to thank also Z. Ajduk, S. Pokorski, and
the organizing committee for the warm hospitality during the workshop.


\begin{figure}
\caption{The $m_{t}^{pole} - \tan\beta$ plane is divided
into five different regions.
Two areas (low- and high-$\tan\beta$ branches) are consistent with perturbative
Yukawa unification ($h_{b} = h_{\tau}$ at $M_{G}$)
and with $0.85m_{b}^{0}(5\,\GeV) < 4.45$ GeV.
Between the two branches the $b$-quark
mass is too high. For a too low (high) $\tan\beta$, $h_{t}$
($h_{b}$) diverges.
The strip where all three (third-family) Yukawa couplings unify intersects
the allowed high-$\tan\beta$ branch and is indicated as well (dash-dot line).
$h_{t} > h_{b}$ above that line.
Corrections to the $h_{t}/h_{b}$ ratio induce a $\sim \pm 5\%$ uncertainty
in the $m_{t}^{pole}$ range that corresponds
to each of the points in the three-Yukawa strip.
$\alpha_{s}(M_{Z})$, $\alpha_{G}$, and the unification scale
used in the calculation are the ones
predicted by the MSSM coupling constant unification.
The upper bound on the $m_{t}^{pole}$  range
suggested by the electroweak data
and $\tan\beta = 1$ are indicated (dashed lines).
$m_{t}^{pole}$   is in GeV.}
\end{figure}

\end{document}